\documentclass[twocolumn, table]{aastex631}
\usepackage{natbib}
\usepackage{hyperref}
\usepackage{rotating}
\usepackage{tabularx}
\usepackage{float}
\usepackage{fontawesome} 

\shorttitle{The Sparkler: High-z Globular Clusters with JWST}
\shortauthors{Mowla \& Iyer et al.}

\newcommand\trs{the Sparkler~}

\begin{document}

\title{The Sparkler: Evolved High-Redshift Globular Clusters Captured by JWST}

\author[0000-0002-8530-9765]{Lamiya Mowla}
\altaffiliation{Equal contribution}
\affiliation{Dunlap Institute for Astronomy and Astrophysics, 50 St. George Street, Toronto, Ontario M5S 3H4, Canada}
\correspondingauthor{Lamiya Mowla and Kartheik Iyer}
\email{lamiya.mowla@utoronto.ca, kartheik.iyer@dunlap.utoronto.ca}

\author[0000-0000-0000-0000]{Kartheik G. Iyer}
\altaffiliation{Equal contribution}
\affiliation{Dunlap Institute for Astronomy and Astrophysics, 50 St. George Street, Toronto, Ontario M5S 3H4, Canada}

\author[0000-0001-8325-1742]{Guillaume Desprez}
\affiliation{
Department of Astronomy \& Physics and Institute for Computational Astrophysics, Saint Mary's University, 923 Robie Street, Halifax, Nova Scotia, B3H 3C3, Canada
}

\author[0000-0001-8489-2349]{Vicente Estrada-Carpenter}
\affiliation{
Department of Astronomy \& Physics and Institute for Computational Astrophysics, Saint Mary's University, 923 Robie Street, Halifax, Nova Scotia, B3H 3C3, Canada
}

\author[0000-0003-3243-9969]{Nicholas S. Martis}
\affil{NRC Herzberg, 5071 West Saanich Rd, Victoria, BC V9E 2E7, Canada}
\affiliation{
Department of Astronomy \& Physics and Institute for Computational Astrophysics, Saint Mary's University, 923 Robie Street, Halifax, Nova Scotia, B3H 3C3, Canada
}

\author{Ga\"el Noirot}
\affiliation{
Department of Astronomy \& Physics and Institute for Computational Astrophysics, Saint Mary's University, 923 Robie Street, Halifax, Nova Scotia, B3H 3C3, Canada
}

\author[0000-0001-8830-2166]{Ghassan T. Sarrouh}
\affiliation{Department of Physics and Astronomy, York University, 4700 Keele St., Toronto, Ontario MJ3 1P3, Canada}

\author[0000-0002-6338-7295]{Victoria Strait}
\affiliation{Cosmic Dawn Center (DAWN), Denmark}
\affiliation{Niels Bohr Institute, University of Copenhagen, Jagtvej 128, DK-2200 Copenhagen N, Denmark}

\author[0000-0003-3983-5438]{Yoshihisa Asada}
\affiliation{Department of Astronomy, Kyoto University, Sakyo-ku, Kyoto 606-8502, Japan}
\affiliation{
Department of Astronomy \& Physics and Institute for Computational Astrophysics, Saint Mary's University, 923 Robie Street, Halifax, Nova Scotia, B3H 3C3, Canada
}

\author[0000-0002-4542-921X]{Roberto G. Abraham}
\affiliation{David A. Dunlap Department of Astronomy and Astrophysics, University of Toronto, 50 St. George Steet, Toronto, Ontario M5S 3H4, Canada}
\affiliation{Dunlap Institute for Astronomy and Astrophysics, 50 St. George Street, Toronto, Ontario M5S 3H4, Canada}

\author[0000-0000-0000-0000]{Gabriel Brammer}
\affiliation{Niels Bohr Institute,
University of Copenhagen,
Blegdamsvej 172100 Copenhagen, Denmark}

\author[0000-0002-7712-7857]{Marcin Sawicki}
\affiliation{
Department of Astronomy \& Physics and Institute for Computational Astrophysics, Saint Mary's University, 923 Robie Street, Halifax, Nova Scotia, B3H 3C3, Canada
}

\author[0000-0002-4201-7367]{Chris J. Willott}
\affil{NRC Herzberg, 5071 West Saanich Rd, Victoria, BC V9E 2E7, Canada}

\author[0000-0001-5984-0395]{Marusa Bradac}
\affiliation{University of Ljubljana, Department of Mathematics and Physics, Jadranska ulica 19, SI-1000 Ljubljana, Slovenia}
\affiliation{Department of Physics and Astronomy, University of California Davis, 1 Shields Avenue, Davis, CA 95616, USA}

\author[0000-0000-0000-0000]{Ren\'e Doyon}
\affiliation{D\'epartement de physique,
Universit\'e de Montr\'eal,
Complexe des Sciences,
PO Box 6128, Centre-ville STN
Montreal QC H3C 3J7,
Canada}

\author[0000-0000-0000-0000]{Kate Gould}
\affiliation{Niels Bohr Institute,
University of Copenhagen,
Blegdamsvej 172100 Copenhagen, Denmark}

\author[0000-0000-0000-0000]{Adam Muzzin}
\affiliation{Department of Physics and Astronomy, York University, 4700 Keele St., Toronto, Ontario MJ3 1P3, Canada}

\author[0000-0000-0000-0000]{Camilla Pacifici}
\affiliation{Space Telescope Science Institute, 3700 San Martin Drive, Baltimore, MD 21218 USA}

\author[0000-0002-5269-6527]{Swara Ravindranath}
\affiliation{Space Telescope Science Institute, 3700 San Martin Drive, Baltimore, MD 21218 USA}

\author[0000-0000-0000-0000]{Johannes Zabl}
\affiliation{
Department of Astronomy \& Physics and Institute for Computational Astrophysics, Saint Mary's University, 923 Robie Street, Halifax, Nova Scotia, B3H 3C3, Canada
}

\begin{abstract}
Using data from JWST, we analyze the compact sources (‘sparkles’) located around a remarkable $z_{\rm spec}=1.378$ galaxy (the `Sparkler’) that is strongly gravitationally lensed by the $z=0.39$ galaxy cluster SMACS J0723.3-7327.
Several of these compact sources can be cross-identified in multiple images, making it clear that they are associated with the host galaxy.
Combining data from JWST’s {\em Near-Infrared Camera} (NIRCam) with archival data from the {\em Hubble Space Telescope} (HST), we perform 0.4--4.4$\mu$m photometry on these objects, finding several of them to be very red and consistent with the colors of quenched, old stellar systems.
Morphological fits confirm that these red sources are spatially unresolved even in strongly magnified JWST/NIRCam images, while JWST/NIRISS spectra show [OIII]5007 emission in the body of the Sparkler but no indication of star formation in the red compact sparkles.
The most natural interpretation of these compact red companions to the Sparkler is that they are evolved globular clusters seen at $z=1.378$.
Applying \textsc{Dense Basis} SED-fitting to the sample, we infer formation redshifts of $z_{form} \sim 7-11$ for these globular cluster candidates, corresponding to ages of $\sim 3.9-4.1$ Gyr at the epoch of observation and a formation time just $\sim$0.5~Gyr after the Big Bang. If confirmed with additional spectroscopy, these red, compact “sparkles” represent the first evolved globular clusters found at high redshift, could be amongst the earliest observed objects to have quenched their star formation in the Universe, and may open a new window into understanding globular cluster formation. Data and code to reproduce our results will be made available at \faGithub\href{https://niriss.github.io/sparkler.html}{http://canucs-jwst.com/sparkler.html}.\\
\end{abstract}


\section{Introduction} \label{sec:intro}

Despite being the subject of very active research for decades \cite[see, e.g., reviews by][]{1979ARA&A..17..241H,1981ARA&A..19..319F, 2006ARA&A..44..193B, forbes2018}, we do not know when, or understand how, globular clusters form. We {\em do} know that most globular clusters in the Milky Way, and those around nearby galaxies, are very old.  The absolute ages of the oldest globular clusters, determined by main sequence fitting and from the ages of the oldest white dwarfs, are about 12.5~Gyr, corresponding to formation redshifts of $z_{\rm form}\sim5$. However, the uncertainties in age estimates are relatively large compared to cosmic age of the Universe at high reshifts, and absolute ages corresponding to $z_{\rm form}\sim3$ (at Cosmic Noon) at the low end, and extending well into the epoch of reionization at the high end ($z_{\rm form}\gg 6$), are plausible \citep{forbes2018}. 

There are two general views on how globular clusters formed. In the first, globular cluster formation is a phenomenon occurring predominantly at very high redshift, with a deep connection to initial galaxy assembly. Ideas along these lines go back to \citet{1968ApJ...154..891P}, who noted that the typical mass of a globular cluster is comparable to the Jeans mass shortly after recombination. In this view, globular clusters are a special phenomenon associated with conditions in the early Universe, and their formation channel is different from that driving present-day star formation. The second view associates globular clusters with young stellar populations seen in nearby starbursting and merging galaxies \citep{Schweizer98, deGrijs2001}. In this case, globular cluster formation might be a natural product of continuous galaxy evolution in systems with high gas fractions, and globular cluster formation would peak at lower redshifts \citep{TrujilloGomez2021}.


We are on the cusp of distinguishing observationally between these two globular cluster formation channels. The {\em JWST} is capable of  observing routinely down to nJy flux levels at wavelengths beyond two microns, and thus of observing globular cluster formation occurring at high redshift \citep{carlberg2002,renzini2017, vanzella17,vanzella2022}. 
In this paper, we use newly-released data from JWST to analyze the nature of the point sources seen around a remarkable multiply-imaged galaxy at $z=1.378$ that we fondly named the `Sparkler'. One image of this galaxy is strongly magnified by a factor of $\sim10-100$ \citep{mahler22, caminha22} by the $z=0.39$ galaxy cluster SMACS J0723.3–7327 (hereafter SMACS0723). Our goal is to determine whether these point sources are (1) globular clusters, (2) super star clusters in the body of the galaxy, or (3) the product of global star-formation in this galaxy being driven by some other mechanism.

Throughout this paper we use AB magnitudes and assume a flat cosmology with $\Omega_m=0.3$, $\Omega_\Lambda=0.7$ and $H_0 = 70$ km s$^{-1}$ Mpc$^{-1}$.

\begin{figure*}[!ht]
\begin{center}
\includegraphics[width=\textwidth]{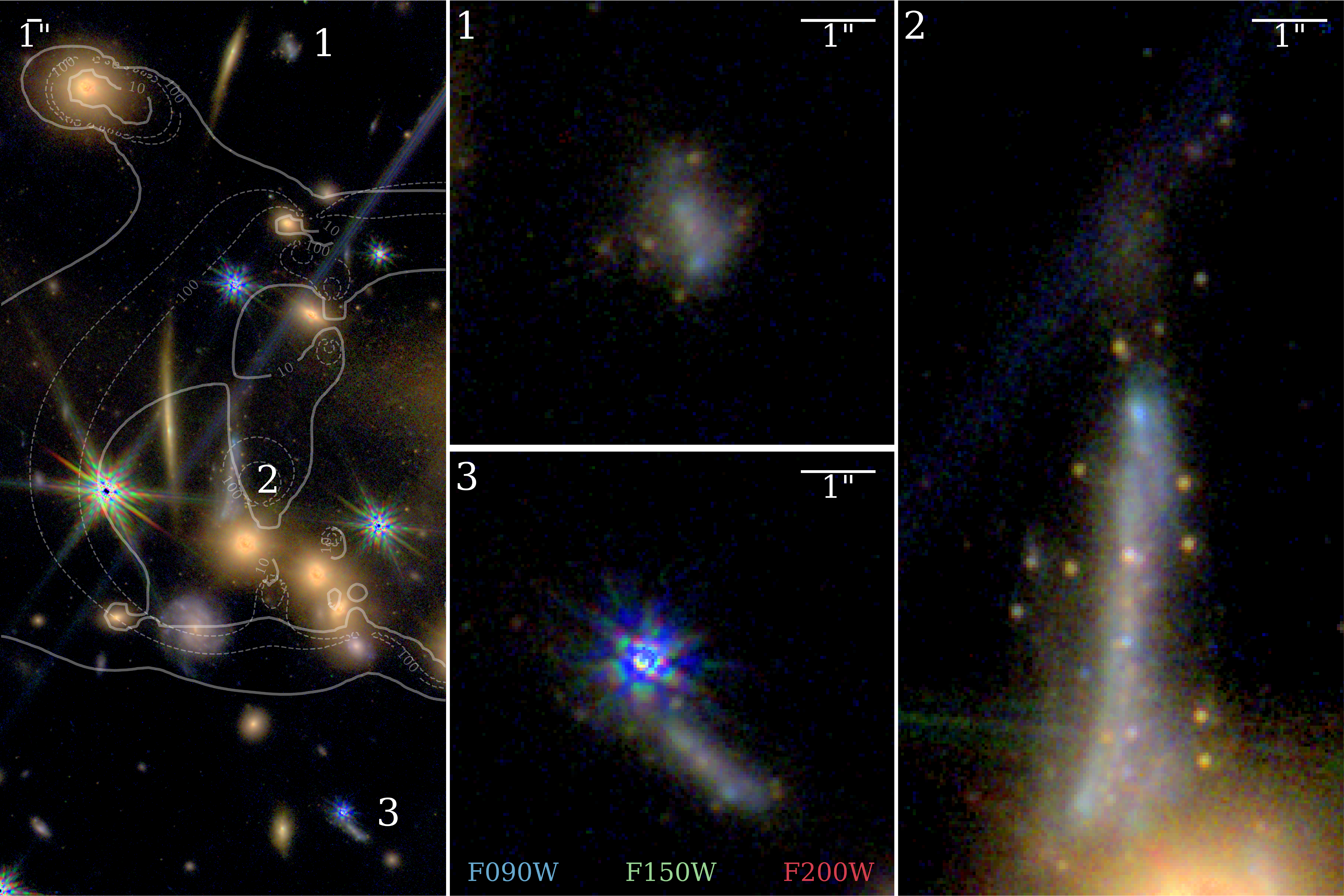}
\end{center}
\caption{Color images of the Sparkler and its environs made by combining F090W, F150W, and F200W images at native spatial resolution. The left panel shows the region around the three images of The Sparkler, with lines of lensing magnification from the \citet[solid curves]{mahler22} and \citet[dashed curves]{caminha22} models overlaid.  Note that regions of very strong magnification ($\mu\sim10-100$) cross image 2 of The Sparkler. The remaining three panels zoom in on the three images of this galaxy. Images are centered on the following positions. Image 1: RA$=$110.83846, Dec$=$-73.45102; Image 2: RA$=$ 110.84051, Dec$=$-73.45487, and Image 3: RA$=$ 110.83614, Dec$=$73.45879.  Note the compact sources, many of them red, surrounding the body of the galaxy; these are most prominent in image 2, but are also discernible in images 1 and 3.}
\label{fig:1}
\end{figure*}

\section{Data}
The imaging and wide field slitless spectroscopy data used for this work are from JWST ERO program 2736 (``Webb's First Deep Field"; \citealt{2022arXiv220713067P}). The galaxy cluster was observed with all four instruments on JWST. Only {\em Near Infrared Camera} (NIRCam; \citealt{rieke2005}) imaging, and {\em Near Infrared Imager and Slitless Spectrograph} (NIRISS; \citealt{doyon2012}) spectroscopy is used in this paper. NIRCam imaging is available in six broad-band filters: F090W, F150W, F200W, F277W, F356W and F444W. Shallow NIRISS wide-field spectroscopy was obtained in the F115W and F200W filters with the two orthogonal low resolution grisms to mitigate contamination \citep{willott22}. Only the F115W grism data is used in this study, because it is the only filter containing a strong emission line, [OIII]$\lambda5007$. These JWST data are supplemented with HST/ACS imaging in F435W and F606W from the RELICS program, drizzled to the same pixel grid \citep{Coe2019}.

We reduced all imaging and slitless spectroscopic data together using the \texttt{Grizli}\footnote{\href{https://github.com/gbrammer/grizli}{https://github.com/gbrammer/grizli}} (\citealp{2021zndo...5012699B}) grism redshift and line analysis software for space-based spectroscopy package. We first obtained uncalibrated ramp exposures from the Mikulski Archive for Space Telescopes (MAST\footnote{\href{https://archive.stsci.edu/}{https://archive.stsci.edu/}}), and ran a modified version of the JWST pipeline stage Detector1, which makes detector-level corrections for, e.g., ramp fitting, cosmic ray rejection (including extra ``snowball" artifact flagging), dark current, and calculates ``rate images". Our modified version of the pipeline also includes a column-average correction for $1/f$-noise. Subsequently, we used the preprocessing routines in \texttt{Grizli} to align all exposures to HST images, subtracted the sky background, and drizzled all images to a common pixel grid with scale 0\farcs04 per pixel. For the NIRCam F090W, F150W and F200W images we created another data product on a 0\farcs02 pixel scale.  The context for the JWST Operational Pipeline (\texttt{CRDS\_CTX}) used for reducing the NIRISS (NIRCam) data was \texttt{jwst\_0932.pmap} (\texttt{jwst\_0916.pmap}). This is a pre-flight version of the NIRCam reference files, so the NIRCam fluxes should be treated with caution.  One consequence of this is that the NIRCam photometric zeropoints calculated from our reductions may be incorrect for in-flight performance, so we used \texttt{EAZY} \citep{2008ApJ...686.1503B} to derive zeropoint offsets consistent with photometric redshift fitting of the full source catalog. Bright cluster galaxies and the intracluster light were modelled and subtracted out using custom code (N.~Martis et al., in preparation). To enable measurement of accurate colors our analysis was done after convolution by a kernel to match the point spread function (PSF) of F444W. 

\begin{figure*}[t]
\begin{center}
\includegraphics[width=\textwidth]{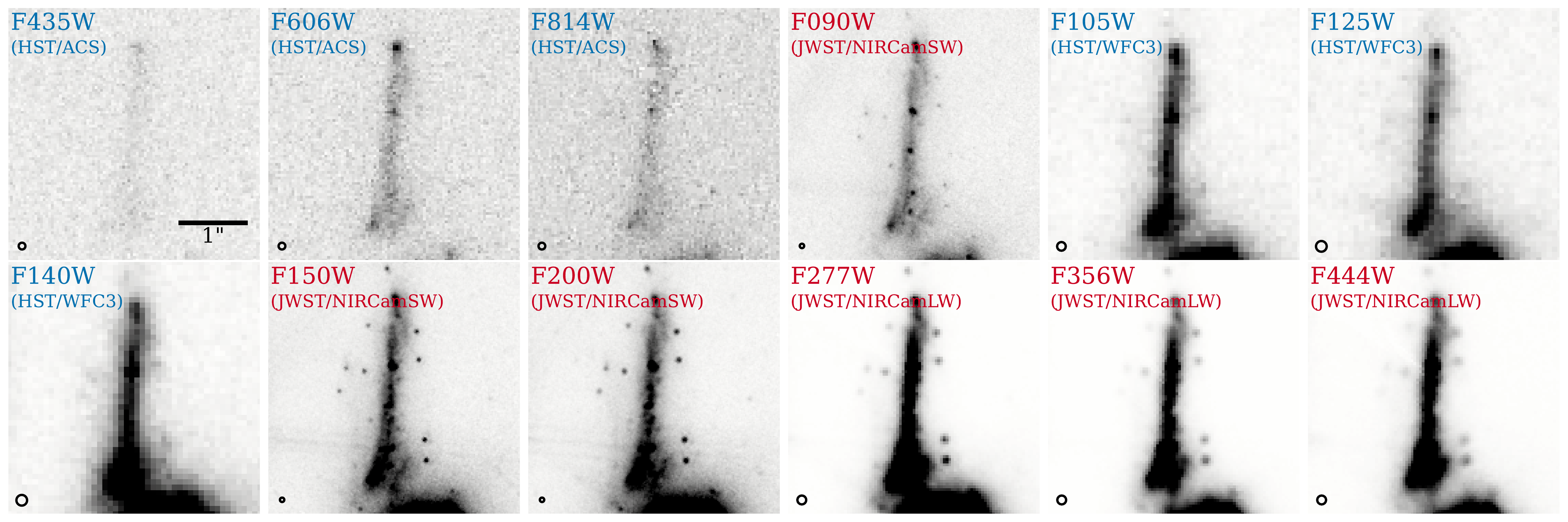}
\end{center}
\caption{Image 2 of the Sparkler from {\em HST/Advanced Camera Survey} (HST/ACS), {\em HST/Wide Field Camera 3} (HST/WFC3), and {\em JWST/Near Infrared Camera} (JWST/NIRCam) Short Wavelength (SW) and Long Wavelength (LW) at observed wavelengths from 0.4--4.4$\mu$m. Circles in the lower-left of each panel show the full width half maximum of the point spread function. Note the exquisite resolution of JWST/NIRCamSW reveals the compact sources surrounding the galaxy, which were not resolved by HST in earlier observations at similar wavelengths.}
\label{fig:2}
\end{figure*}

Figure \ref{fig:1} shows images of the Sparkler. Coordinates for the three images of the background galaxy are presented in the caption accompanying the figure. The Sparkler was first identified as multiply-imaged in HST imaging combined with ESO MUSE integral-field spectroscopy that shows all three images having [OII]$\lambda3727$ emission \citep{golubchik22}. We adopt the spectroscopic redshift of $z=1.378 \pm 0.001$ from the MUSE [OII] line (\citealt{mahler22, caminha22, golubchik22}). The magnifications of the three images (labeled as 1, 2, and 3 in Figure~\ref{fig:1}) in the lensing model of \citet[their IDs 2.1, 2.2, and 2.3]{mahler22}
are 3.6$\pm$0.1, 14.9$\pm$0.8 and 3.0$\pm$0.1, respectively. In the lensing model of \citet[their IDs 3a, 3b, and 3c]{caminha22} the magnifications are significantly higher: $9.2^{+1.3}_{-1.2}, 103^{+153}_{-47},$ and $6.1^{+0.7}_{-0.7}$, respectively. Based on measured flux ratios between the three images we consider the \citet{caminha22} model to better fit the properties of this galaxy. As shown in Figure~\ref{fig:1}, there may be critical curves and/or high magnification contours crossing image 2 (magnification 5--10 in the \citealt{mahler22} model and magnification 30--100+ in the \citealt{caminha22} model),  suggesting strong differential magnification in the image. 
Figure~\ref{fig:2} shows a multi-band montage of Image 2 of the Sparkler, using data from HST/ACS, HST/WFC3, and JWST/NIRCam short- and long-wavelength cameras at observed wavelengths spanning 0.4-4.4$\mu$m. 
Circles in the lower-left of each panel show the full width half maximum of the point spread function. The exquisite resolution of JWST/NIRCam SW best reveals the compact sources surrounding the galaxy, which were not resolved by HST in earlier observations even 
at similar wavelengths.\\



\begin{figure*}
    \centering
    \includegraphics[width=\textwidth]{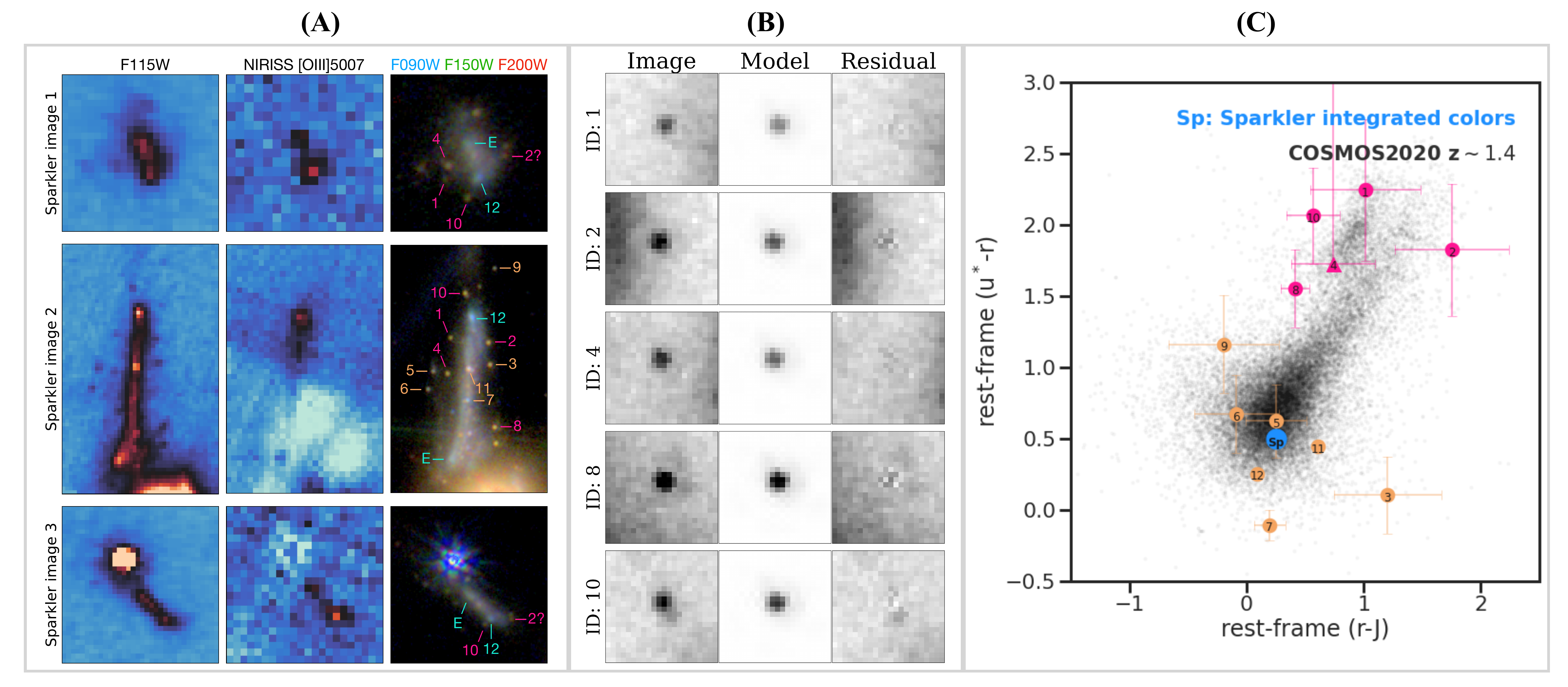}
    \caption{\textbf{(A): The globular cluster candidates are associated with the main galaxy.} F115W images (left column), [OIII]$\lambda$5007 emission line maps derived from the NIRISS grism data in the F115W band (middle column), and NIRCam color composite images (right column). Sparkle IDs are shown for Image 2, with tentative counterparts identified in Images 1 and 3. The lower part of the [OIII] map of Image 2 suffers from significant contamination. [OIII] emission is a classic signature of ongoing star formation; here, it is present in the star-forming regions of the host galaxy, but its absence at the locations of the globular cluster candidates supports the hypothesis that at the epoch of observation these are quiescent systems. 
    \textbf{(B): The globular cluster candidates are unresolved.} Fits to the globular cluster candidates with point sources on the 0\farcs02 F150W images using GALFIT \citep{Peng2010} show that the residuals are consistent with noise. 
    \textbf{(C): The globular cluster candidates have colors of quenched stellar systems.} 
    $urJ$ colors 
    (measured \emph{directly} from F090W, F200W, and the average of F277W and F356W fluxes) compared with $z\sim 1.4$ galaxies in the COSMOS2020 catalog \citep{weaver22}: the integrated colors of the Sparkler galaxy (blue circle labeled Sp) are in the star-forming blue cloud, as are our
    other point sources ({\color{orange} \textbf{orange}}), but the globular cluster candidates ({\color{magenta} \textbf{pink}}) have u$^*$-r$>$1.5 and are consistent with the colors of quenched systems.
    }
    \label{fig:OIII}
\end{figure*}

\begin{figure*}
    \centering
    \includegraphics[width=\textwidth]{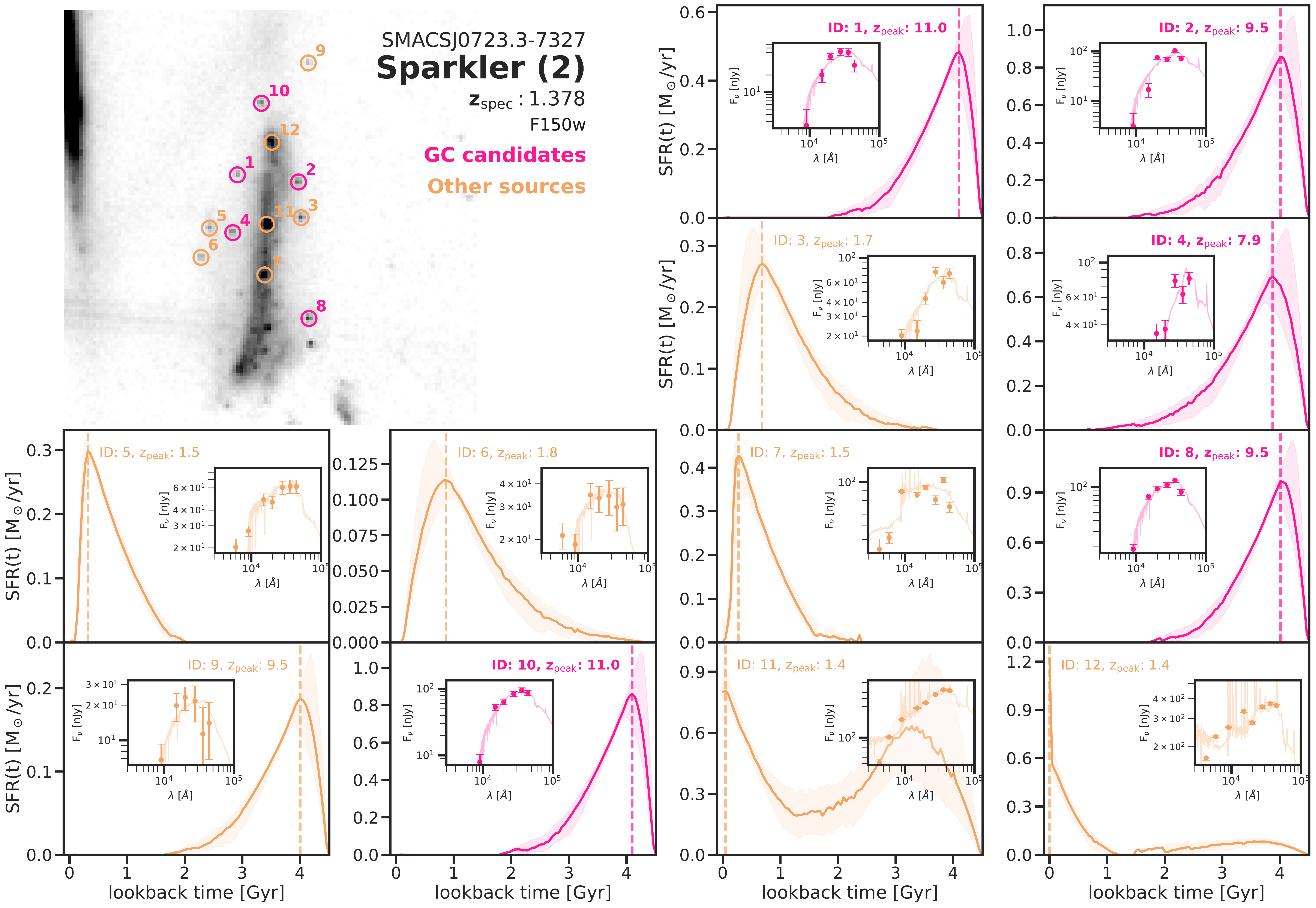}
    \caption{Non-parametric SFHs derived from fitting the photometric SEDs of the individual sparkles. {\color{magenta} \textbf{Pink}} points and curves show the locations and colors (\textit{top left}), SFHs (\textit{marked panels}) and SED fits (\textit{inset panels}) of the individual globular cluster candidates, while {\color{orange} \textbf{orange}} is used to show fits and SFHs for objects that are extended sources, heavily contaminated by light from the galaxy, nearby objects or ICL, or in the body of the main galaxy. Even though object 9 is consistent with an early SFH, we exclude it as a globular cluster candidate due to low SNR and possible contamination by a nearby diffraction spike.  SEDs are shown in $F_\nu$ units, with the spectra corresponding to the best-fit model from \textsc{Dense Basis}. SFR values are not corrected for lensing magnification, which could make them $\sim$10-100 times smaller. $z_{\rm peak}$ corresponds to the redshift at which the posterior SFH peaks in SFR. 
    Overall, the globular cluster candidates show SFHs consistent with very early epochs of star formation ranging over $7<z<11$. 
    }
    \label{fig:sparkle_sfhs}
\end{figure*}

\section{Methods}

In this letter we focus our attention on twelve compact candidates in and around the Sparkler. In this preliminary exploration, we selected candidates by eye, focusing mainly on compact objects (`sparkles') in uncontaminated regions of the image.  A few compact sources in the galaxy itself were also added to our sample to allow us to compare objects in the body of the Sparkler to objects in the periphery of the galaxy. Objects were selected using the very deep 0\farcs02 pixel scale F150W image, and were chosen to be broadly representative of the compact sources in this system. As described below, 2D modeling confirms that the objects chosen are unresolved. We emphasize that the objects analyzed in this letter are not a complete sample. Construction of a complete sample will require detailed background subtraction and foreground galaxy modelling, which is deferred to a future paper.

\subsection{Aperture Photometry} 

Photometry is challenging in crowded fields, and in the case of the Sparkler the challenges are compounded by contamination from the host galaxy and from other nearby sources. This contamination can significantly alter the shape of the SED of the individual compact sources. In a future paper we will present a full catalog of compact sources around the Sparkler that attempts to account for these effects by subtracting contamination models and using PSF photometry. For simplicity and robustness, in the present paper we used aperture photometry, as this technique is relatively insensitive to variations in the local background.  Photometry was done using images that i. are on 0\farcs04 pixel scale, ii. have bright cluster galaxy and ICL-subtracted, and iii. are F444W PSF-convolved F435W, F606W, F090W, F150W, F200W, F277W, F356W, F444W images. 

Using \texttt{photutils} \citep{photutils21}, circular apertures with radii of 0\farcs12, 0\farcs16 and 0\farcs20 were defined using the centroided positions of the twelve sparkles in the F150W image. An annulus starting at the edge of the aperture and with width 0\farcs08 was used to estimate the median local background, which was subtracted from the aperture flux. Aperture correction was applied by multiplying with the F444W PSF growth curve. To determine contamination corrections, we injected simulated point sources of various fluxes around the galaxy to determine how well our procedure recovered the intrinsic total flux of the compact sources. We found that the precision of the photometry varied widely across the different filters,  environments, and intrinsic brightness of the sources, but that these variations could be quantified by simulations. For every sparkle, we identified a location proximate to it in which we injected simulated point sources to model the measurement accuracy. For a sparkle at a given wavelength, we injected 20 point sources of total flux varying between 0.1 and 10 times the measured flux of the source and measured their fluxes using the same techniques used to analyze the original sources. We then fit the intrinsic flux as a function of the measured flux with a second-order polynomial, which we used to determine local aperture corrections. This process was repeated across 20 different locations around the galaxy to estimate the uncertainty in flux measurement. We selected the 0\farcs20 aperture for our final photometry as the corrected flux recovered $>99\%$ of intrinsic flux across all environments. The procedure was performed for all twelve sparkles in all eight filters to construct the final SED of the sources. For sources that are undetected, we assigned an upper limit of three times the noise of the image. 

\subsection{SED fitting and estimating physical properties}

Spectral Energy Distributions (SEDs) derived from our aperture photometry were analyzed using the \textsc{Dense Basis} method\footnote{\href{https://dense-basis.readthedocs.io/}{https://dense-basis.readthedocs.io/}} \citep{iyer17, iyer19} to determine non-parametric star-formation histories (SFHs), masses, ages, metallicities and dust extinction values for our compact sources.  The \textsc{Dense Basis} fits were run with a single t$_{50}$ parameter, following the prescription in \citet{iyer19}, with the full methodology and validation tests presented in \citet{iyer18, iyer19} and \citet{olsen21}. The primary advantages of using non-parametric SFHs is that they allow us to account for multiple stellar populations, robustly derive SFH-related quantities including masses, SFRs and ages, and allow us to set explicit priors in SFH space to prevent outshining due to younger stellar populations that could otherwise bias estimates of these properties \citep{iyer17,leja19,lower20}. 

However, \textsc{Dense Basis}, by design, implements correlated star formation rates over time, to better encode the effects of physical processes in galaxies that regulate star formation and to better recover complex SFHs containing multiple stellar populations \citep{iyer19}. The formalism smooths out star formation histories that are instantaneous pulses, and has an age resolution of about 0.5 Gyr. We therefore also  undertook SED fits based on simple luminosity evolution of simple stellar populations (SSPs). As will be seen below, in several cases the \textsc{Dense Basis} fit results return SFHs that are as close to instantaneous pulses as the method allows. In such cases, SSP fits may give comparably good results with fewer assumptions. SSP fits also have the benefit of returning unambiguously-defined ages. Since the \textsc{Dense Basis} fits provide a full SFH posterior, we will define the `age' from these fits to be the time at which the SFR peaks ($t_{\rm peak}$). Using validation tests fitting synthetic SSP sources injected into the field and mock photometry with similar noise properties to the observed sources, we find that this can robustly recover the age of the corresponding SSP within uncertainties, finding a bias and scatter of $(\mu, \sigma)_{t_{50}} \equiv (0.15,1.00)$~Gyr, $(\mu, \sigma)_{t_{\rm peak}} \equiv (0.13,0.86)$~Gyr and $(\mu, \sigma)_{age_{\rm SSP}} \equiv (-0.20,0.86)$~Gyr for the three metrics tested.

\subsection{Grism extraction and fitting}

Before extracting individual NIRISS spectra, we constructed a contamination model of the entire field using \texttt{Grizli}. We modeled sources at both grism orientations. This model was built using a segmentation map and photometric catalog created with SEP (\citealt{Barbary2016}, \citealt{1996A&AS..117..393B}). We initially assumed a flat spectrum, normalized by the flux in the photometric catalog, in our models. Successive higher-order polynomials were then fit to each source, iteratively, until the residuals in the global contamination model were negligible.


After the spectral modelling of the full field for contamination removal, we then extracted the 2D grism cutouts of the three images of the Sparkler and fitted their spectra using the \texttt{Grizli} redshift-fitting routine with a set of \texttt{FSPS} and emission line templates. \texttt{Grizli} forward model the 1D spectral template set to the 2D grism frames based on the source morphologies in the direct imaging. Based on the grism data alone, \texttt{Grizli} identified multiple redshift solutions for the Sparkler including a solution at $z=1.38$ based on the identification of [OIII]$\lambda$5007 at 1.2$\mu$m in the F115W grism data. This is consistent with the identification of the complementary OII line previously reported in the MUSE data, and securely confirms the spectroscopic redshift of the source as $z=1.38$. As a product of the fitting, emission-line maps of the [OIII]$\lambda$5007 line were created for the three images of the Sparkler.

\section{Results}

The fluxes and associated uncertainties for the twelve compact sources (`sparkles') in and around the Sparkler are presented in Table \ref{tab:sparkle_properties} and their positions are identified on Image 2 of the Sparkler in the middle row of panel (A) in Figure~\ref{fig:OIII}. Panel (B) of this figure shows point source fits (using GALFIT; \citealt{Peng2010}) to several sparkles in our sample. Residuals from the fits are negligible, confirming the original visual impression that these compact sources are unresolved. Panel (C) in Figure~\ref{fig:OIII} shows the colors of the individual sparkles in the rest-frame $urJ$ color-color space (measured {\emph {directly}} from F090W, F200W, and the average of F277W and F356W fluxes), overplotted on the distribution of $z\sim 1.4$ galaxies from the COSMOS2020 catalog \citep{weaver22}. The body of the Sparkler galaxy (blue point) is in the star-forming blue cloud, as are 7 of 12 of our sparkles (orange points). However, five of the sparkles have red colors ($u^*-r>1.5$) consistent with those of quiescent systems (the so-called red cloud).  Panel (B) in Figure~\ref{fig:OIII} shows two-dimensional fits of the point-spread function to these reddest five sources (obtained using GALFIT; \citealt{Peng2010}). Residuals from the fits are negligible, confirming the original visual impression that these compact red sources are unresolved.  These five red, unresolved  objects will constitute our sample of globular cluster candidates throughout this paper, and are color-coded in pink in all figures in this paper. 
    
SEDs and derived SFHs inferred from our modelling are shown in Figure~\ref{fig:sparkle_sfhs}. The physical properties corresponding to the models shown in this figure are also given in Table \ref{tab:sparkle_properties}. The table contains effective ages of the globular cluster candidates from both \textsc{Dense Basis} and SSP fitting methods, which generally agree within the uncertainties. Of the objects under consideration, six (IDs 1,2,4,8,9,10) are consistent with SFHs that peaked at early formation times. Note that we do not include Object 9 in our list of globular cluster candidates because of its low SNR, coupled with possible contamination from the nearby diffraction spike and extended tail visible in Figure \ref{fig:1}. Objects 11 and 12, which are in the bulk of the galaxy, show recent star formation, consistent with the [OIII]$\lambda$5007 emission in Figure~\ref{fig:OIII}.

Panel (A) of Figure~\ref{fig:OIII} shows the emission line maps at the redshifted wavelength of [OIII]$\lambda$5007 for all three images of the Sparkler.  Individual columns show the direct, F115W image (the broadband filter within which the redshifted [OIII] emission lies), and a NIRCam F090W, F150W, and F200W  color composite for each Sparkler image. There is clear evidence of [OIII]$\lambda$5007 emission in all three images, which we interpret as related to star-formation activity in the Sparkler.  Note that the line emission is spatially co-located with the two blue regions in the color composite, consistent with this interpretation.  Most importantly, there is no evidence of line emission at the locations of those sparkles that we have previously identified as globular cluster candidates (IDs 1, 2, 4, 8, and 10), and this adds confidence to our conclusion that these objects consist of old stellar populations and are devoid of ongoing star formation. 

Much can be learned from inter-comparing the images shown in Panel (A) of Figure~\ref{fig:OIII}, and in particular, from comparing the properties of sparkles we identified in Image 2 with their counterparts in Images 1 and 3. We leave such analysis, as well as the construction of a full lens model of the system, to future papers; for now, we simply highlight a few tentatively matched features in the third column of this panel, focusing on the globular cluster candidates (pink labels) and the two most prominent star-forming regions (cyan labels).

We close this section with some preliminary discussion of the mass of the Sparkler. Fits to the integrated photometry of images 1 and 3 using \textsc{Dense Basis} recover log stellar masses of ${9.67}^{+0.08}_{-0.09}$M$_\odot$ and ${9.51}^{+0.08}_{-0.08}$M$_\odot$ respectively for the host galaxy (uncorrected for magnification), and star formation histories that show a recent rise over the last $\sim$ Gyr. We do not fit image 2 due to the strong differential magnification. Assuming magnifications of $\sim 5$ for these images (much lower than for image 2), the stellar mass of Sparkler would be around $10^9~{\rm M}_\odot$ , which is similar to that of the Large Magellanic Cloud \citep{erkal19}, which has $\sim40$ globular clusters \citep{bennet22}.

\section{Discussion}

We are at the earliest stages of understanding how best to calibrate data from the in-orbit JWST, so SED modeling is best approached with a degree of caution. For this reason, we emphasize that our most important conclusions spring from observations that are independent of detailed SED modeling. Firstly, many of the compact sources in and around the Sparkler are unresolved (panel B of Figure~\ref{fig:OIII}) and several can be cross-identified in multiple images (Figure~\ref{fig:1} and panel A in Figure~\ref{fig:OIII}), so they are clearly associated with the host galaxy, placing them at $z=1.378$. The colors of these systems are consistent with the expected positions of quiescent sources at $z=1.378$ on a rest-frame $urJ$ diagram (panel C of Figure~\ref{fig:OIII}). Independently of any modeling, these facts suggest an identification of the red sparkles with evolved globular clusters.

Going further than this requires modeling. At face value, the reddest compact clumps (five of the twelve in Table \ref{tab:sparkle_properties} and Figure \ref{fig:sparkle_sfhs}) surrounding \trs show SFHs consistent with simple stellar populations formed at very high redshifts ($z \gtrsim 9$). Another two objects, mainly in the bulk of the galaxy, show SFHs consistent with younger ($\sim0.03-0.3$ Gyr) stellar populations.

The quiescent nature of the reddest point sources in and around \trs effectively rules out the possibility that they are active star formation complexes of the kind seen in many $1 < z < 3$ galaxies, such as those associated with dynamical instabilities in gas-rich turbulent disks \citep{Genzel2006,Schreiber2006}. 
A number of studies examining clumps in high-redshift systems with strong gravitational lensing have been able to explore the clump size distribution at physical spatial resolutions below 100~pc (e.g., \citealt{livermore12,wuyts14,livermore15, johnson17,welch2022}). These report a broad range of sizes (50 pc -- 1 kpc), but because of the high magnification of the Sparkler, most such clumps would be expected to be resolved by the JWST data we study. As already noted, pioneering work by \citet{johnson17} and \citet{vanzella17} suggests that HST observations of strongly lensed active star-formation complexes in galaxies at $2<z<6$ may already have captured the earliest phases of globular cluster formation. More recent work on lensed $z\sim6$ galaxies has revealed even smaller complexes, e.g.\ in the Sunrise Arc \citep{welch22}. This work is exciting, but the association of young massive clusters at high redshift with proto-globular clusters remains indirect, and the future evolution of these star formation complexes is unclear. 

The most interesting interpretation of the clumps in and around \trs is that the bulk of them are evolved (maximally old, given the 4.6 Gyr age of Universe at the epoch of observation) globular clusters. If this interpretation is correct, JWST observations of quiescent, evolved globular clusters around $z\sim1.5$ galaxies can be used to explore the formation history of globular clusters in a manner that is complementary to searching directly for the earliest stages of globular cluster formation (e.g., by examining young massive star-formation complexes at $z\sim6$ and higher). Young star formation complexes may, or may not, evolve eventually into globular clusters, but there can be little doubt about the identity of an isolated and quiescent compact system if its mass is around $10^6$ M$_\sun$ and its scale length is a few parsec. JWST observations of evolved globular clusters at $z\sim1.5$ are also complementary to exploration of the ages of {\em local} globular clusters, as models fit to local globular clusters cannot distinguish between old and very old systems. For example, distinguishing between an $\sim 11.5$ Gyr old stellar population that formed at $z=3$ and a 13.2 Gyr old stellar population that formed at $z=9$ is not possible with current models and data, because they are degenerate with respect to a number of physical parameters \citep{ocvirk06, conroy09_fsps, conroy10_fsps}. JWST observations of {\em evolved} globular clusters, seen when the Universe was about one third of its present age, provide an opportunity for progress by `meeting in the middle', because population synthesis models of integrated starlight from simple stellar populations {\em can} distinguish rather easily between the ages of young-intermediate stellar populations. This is because intermediate-mass stars with very distinctive photospheric properties are present at these ages. At $z=1.378$, the lookback time to \trs is 9.1 Gyr, and the age of the Universe at that epoch is 4.6 Gyr. Distinguishing between $z=3$ and $z=9$ formation epochs for the globular cluster system corresponds to distinguishing between 2.4 Gyr- and 4.1 Gyr-old populations, which is relatively straightforward for population synthesis models in the JWST bands. In the case of the Sparkler, the striking conclusion is that at least 4 of its globular clusters have likely formed at $z>9$. 

Our identification of the `sparkles' in Figure~1 with evolved globular clusters relies on an assumption of very strong magnification of the Sparkler. Strong magnification occurs only in narrow regions near lensing caustics, so there are strong magnification gradients in the source plane. This makes it difficult to invert lens models to compute accurate luminosity functions for the putative globular cluster population. Based on Figure \ref{fig:1}, we assume the overall magnification of the system is large (at least a factor of 15), but handling the strong magnification gradients across the local environment of \trs is beyond the scope of this paper. 
Assuming magnifications of 10--100, the stellar masses of these point sources fall in the range $\sim 10^6-10^{7}$M$_\odot$, which is plausible for metal-poor globular clusters seen at ages of around 4 Gyr, although most lie at the high end of the local globular cluster mass range. Since critical curve may be running through the system, we emphasize again that the magnification (and hence the masses) of the clusters is very uncertain. 
 
If lens models can be determined with the accuracy needed to compute source plane luminosity functions and mass distributions, then the Sparkler may place interesting constraints on globular cluster dissolution. Physical processes slowly dissolve globular clusters, and luminosity evolution is significant, so distant globular clusters are expected to be both more massive and more luminous than their local counterparts. The most relevant physical processes are stellar evolution coupled with relaxation and tidal effects, and in some models significant mass loss is expected. For example, with a standard Kroupa IMF \citep{Kroupa2001} about 30\% of the mass of a star cluster is expected to be lost due to stellar evolution alone in the first few Gyr \citep{Baumgardt2003}, and this fraction is much higher for top-heavy IMFs. Dynamical processes would compound this loss, though dynamical processes are likely to be most significant for lower mass clusters \citep{2006astro.ph..5125B}. In any case, unless globular cluster dissolution processes are operating far more quickly than expected, very high magnifications are certainly needed to explain the point sources surrounding \trs as globular clusters.

\section{Conclusions}

{\em In situ} investigations of evolved globular cluster systems at $z\sim1.5$ present us with a golden opportunity to probe the initial formation epoch of globular clusters with a precision unobtainable from studying local systems. Magnified red point sources seen at this epoch are old enough to be unambiguously identified as globular clusters, but young enough that their ages can be determined quite reliably. We applied this idea to JWST and HST observations of a $z=1.378$ galaxy (which we refer to as the Sparkler), which is strongly lensed by the $z=0.39$ galaxy cluster SMACS J0723.3–7327. At least five of the twelve compact sources in and around the Sparkler are unresolved and red, and the most likely interpretation of these is that they are {\em evolved} globular clusters seen at $z=1.378$. By modeling the colors and spectra of these compact sources with the \textsc{Dense Basis} method, four (33\%) are found to be consistent with simple stellar populations forming at $z > 9$, {\em i.e.}, in the first 0.5~Gyr of cosmic history and more than 13~Gyr before the present epoch. If these ages are confirmed, at least some globular clusters appear to have formed contemporaneously with the large-scale reionization of the intergalactic medium, hinting at a deep connection between globular cluster formation and the initial phases of galaxy assembly. Data and code to reproduce our results will be made available at \faGithub\href{https://niriss.github.io/sparkler.html}{http://canucs-jwst.com/sparkler.html}.

\begin{table*}
    \caption{Photometric properties and derived physical parameters for the compact point sources estimated from \textsc{dense basis} and the SSP fitter. Reported uncertainties represent the $16^{th}-84^{th}$ percentiles of the posterior distribution for each quantity.}
    \centering
    \begin{tabular}{c | c c c c c c c c c c c c}
    \hline \hline
       ID & 1 & 2 & 3 & 4 & 5 & 6 & 7 & 8 & 9 & 10 & 11 & 12 \\
\hline
Class$^{a}$ & GC & GC & C & GC & E & E & B & GC & C & GC & B & B \\
F$_\nu$ [nJy; F435W] & -- & -- & -- & -- & -- & -- & $13.98$ & $33.51$ & -- & -- & $38.17$ & $155.42$ \\
$\delta$F$_\nu$ [nJy; F435W] & -- & -- & -- & -- & -- & -- & $\pm 4.57$ & $\pm 4.72$ & -- & -- & $\pm 4.57$ & $\pm 4.63$ \\
F$_\nu$ [nJy; F606W] & -- & -- & -- & -- & $20.28$ & $20.98$ & $21.20$ & -- & -- & -- & $102.37$ & $231.08$ \\
$\delta$F$_\nu$ [nJy; F606W] & -- & -- & -- & -- & $\pm 3.24$ & $\pm 3.28$ & $\pm 3.16$ & -- & -- & -- & $\pm 3.15$ & $\pm 3.08$ \\
F$_\nu$ [nJy; F814W] & -- & -- & -- & -- & $52.41$ & -- & $46.52$ & -- & $67.55$ & -- & $126.44$ & $258.67$ \\
$\delta$F$_\nu$ [nJy; F814W] & -- & -- & -- & -- & $\pm 4.49$ & -- & $\pm 4.54$ & -- & $\pm 4.58$ & -- & $\pm 4.62$ & $\pm 4.54$ \\
F$_\nu$ [nJy; F090W] & $2.40$ & $3.00$ & $19.05$ & -- & $25.70$ & $17.66$ & $72.54$ & $17.40$ & $6.40$ & $7.45$ & $178.21$ & $249.38$ \\
$\delta$F$_\nu$ [nJy; F090W] & $\pm 2.27$ & $\pm 2.26$ & $\pm 2.34$ & -- & $\pm 2.41$ & $\pm 2.42$ & $\pm 1.78$ & $\pm 2.30$ & $\pm 2.35$ & $\pm 2.34$ & $\pm 1.77$ & $\pm 1.87$ \\
F$_\nu$ [nJy; F150W] & $20.12$ & $17.11$ & $22.25$ & $35.21$ & $48.47$ & $34.82$ & $69.74$ & $77.23$ & $19.80$ & $52.99$ & $285.08$ & $334.33$ \\
$\delta$F$_\nu$ [nJy; F150W] & $\pm 5.32$ & $\pm 5.33$ & $\pm 5.28$ & $\pm 5.28$ & $\pm 5.28$ & $\pm 5.28$ & $\pm 5.35$ & $\pm 5.29$ & $\pm 5.25$ & $\pm 5.29$ & $\pm 5.35$ & $\pm 5.32$ \\
F$_\nu$ [nJy; F200W] & $41.05$ & $72.04$ & $41.53$ & $35.97$ & $44.43$ & $32.12$ & $82.67$ & $91.62$ & $22.36$ & $60.53$ & $328.86$ & $271.68$ \\
$\delta$F$_\nu$ [nJy; F200W] & $\pm 5.03$ & $\pm 5.04$ & $\pm 5.09$ & $\pm 5.12$ & $\pm 5.09$ & $\pm 5.14$ & $\pm 5.07$ & $\pm 5.08$ & $\pm 5.11$ & $\pm 5.08$ & $\pm 5.07$ & $\pm 5.04$ \\
F$_\nu$ [nJy; F277W] & $51.28$ & $67.68$ & $73.76$ & $75.72$ & $60.01$ & $33.99$ & $60.36$ & $104.96$ & $21.52$ & $82.81$ & $455.34$ & $351.27$ \\
$\delta$F$_\nu$ [nJy; F277W] & $\pm 7.21$ & $\pm 7.19$ & $\pm 7.24$ & $\pm 7.23$ & $\pm 7.24$ & $\pm 7.34$ & $\pm 7.10$ & $\pm 7.25$ & $\pm 7.33$ & $\pm 7.23$ & $\pm 7.04$ & $\pm 7.06$ \\
F$_\nu$ [nJy; F356W] & $46.60$ & $95.96$ & $55.93$ & $58.01$ & $57.21$ & $27.74$ & $98.49$ & $111.25$ & $10.54$ & $88.50$ & $502.32$ & $345.09$ \\
$\delta$F$_\nu$ [nJy; F356W] & $\pm 7.20$ & $\pm 7.21$ & $\pm 7.17$ & $\pm 7.21$ & $\pm 7.21$ & $\pm 7.16$ & $\pm 7.03$ & $\pm 7.20$ & $\pm 7.14$ & $\pm 7.21$ & $\pm 6.44$ & $\pm 6.98$ \\
F$_\nu$ [nJy; F444W] & $25.70$ & $61.42$ & $62.45$ & $68.21$ & $53.26$ & $26.67$ & $43.28$ & $75.12$ & $12.08$ & $74.95$ & $452.73$ & $311.93$ \\
$\delta$F$_\nu$ [nJy; F444W] & $\pm 6.23$ & $\pm 6.16$ & $\pm 6.21$ & $\pm 6.22$ & $\pm 6.62$ & $\pm 6.13$ & $\pm 6.09$ & $\pm 6.21$ & $\pm 6.17$ & $\pm 6.22$ & $\pm 6.10$ & $\pm 6.07$ \\
\hline
$\log$ M$_{*,50}^b$ [M$_\odot$] & 8.26 & 8.57 & 8.42 & 8.57 & 8.34 & 8.15 & 8.20 & 8.68 & 7.96 & 8.58 & 9.09 & 8.41 \\
$\log$ M$_{*,16}^b$ [M$_\odot$] & 8.15 & 8.48 & 8.32 & 8.48 & 8.24 & 8.01 & 7.72 & 8.60 & 7.82 & 8.49 & 9.01 & 8.33 \\
$\log$ M$_{*,84}^b$ [M$_\odot$] & 8.38 & 8.67 & 8.52 & 8.68 & 8.45 & 8.27 & 8.34 & 8.77 & 8.10 & 8.67 & 9.14 & 8.49 \\
$\log$ sSFR$_{*,50}$ [yr$^{-1}$] & -12.05 & -12.25 & -12.05 & -12.15 & -11.55 & -11.05 & -8.95 & -12.45 & -11.75 & -12.25 & -9.25 & -8.35 \\
$\log$ sSFR$_{*,16}$ [yr$^{-1}$] & -13.25 & -13.35 & -13.25 & -13.35 & -13.15 & -12.95 & -9.75 & -13.35 & -13.15 & -13.35 & -9.35 & -8.45 \\
$\log$ sSFR$_{*,84}$ [yr$^{-1}$] & -10.85 & -11.05 & -10.85 & -10.95 & -10.05 & -9.55 & -8.05 & -11.35 & -10.35 & -11.15 & -9.05 & -8.25 \\
t$_{\rm peak, 50}$ [Gyr] & 4.10 & 4.01 & 0.68 & 3.87 & 0.32 & 0.87 & 0.27 & 4.01 & 4.01 & 4.10 & 0.05 & 0.00 \\
t$_{\rm peak, 16}$ [Gyr] & 1.50 & 1.50 & 0.00 & 1.60 & 0.00 & 0.80 & 0.00 & 1.60 & 1.00 & 1.50 & 0.00 & 0.00 \\
t$_{\rm peak, 84}$ [Gyr] & 4.51 & 4.51 & 1.30 & 0.50 & 1.70 & 1.80 & 0.25 & 4.51 & 4.51 & 4.51 & 3.55 & 0.00 \\
A$_{V,50}^b$ [mag] & 0.30 & 0.48 & 0.28 & 1.28 & 0.15 & 0.16 & 0.18 & 0.23 & 0.22 & 0.34 & 0.27 & 0.03 \\
A$_{V,16}^b$ [mag] & 0.08 & 0.16 & 0.07 & 0.78 & 0.04 & 0.04 & 0.03 & 0.06 & 0.06 & 0.10 & 0.20 & 0.01 \\
A$_{V,84}^b$ [mag] & 0.68 & 0.91 & 0.62 & 1.82 & 0.39 & 0.47 & 0.49 & 0.49 & 0.55 & 0.71 & 0.35 & 0.05 \\
log Z$_{50}$/Z$_\odot$ & -0.51 & -0.33 & -0.47 & -0.11 & -1.04 & -1.02 & -0.82 & -0.72 & -0.72 & -0.36 & 0.21 & 0.11 \\
log Z$_{16}$/Z$_\odot$ & -1.09 & -0.88 & -1.05 & -0.67 & -1.37 & -1.37 & -1.43 & -1.13 & -1.25 & -0.89 & 0.14 & 0.09 \\
log Z$_{84}$/Z$_\odot$ & -0.02 & 0.05 & 0.01 & 0.16 & -0.45 & -0.41 & -0.54 & -0.24 & -0.12 & 0.03 & 0.24 & 0.15 \\
\hline
$\log$ M$_{*,\rm SSP}$ [M$_\odot$] & 8.66 & 8.87 & 8.41 & 8.98 & 8.49 & 8.23 & 8.45 & 9.06 & 8.34 & 8.97 & 8.60 & 8.32 \\
age$_{\rm SSP}$ [Gyr] & 4.37 & 3.16 & 1.26 & 4.50 & 0.56 & 0.71 & 0.39 & 4.47 & 4.47 & 4.50 & 0.04 & 0.03 \\
log Z/Z$_{\odot,{\rm SSP}}$ & -0.50 & -0.26 & -0.55 & 0.00 & -1.47 & -1.50 & -1.50 & -1.00 & -1.50 & -0.50 & 0.00 & 0.00 \\
A$_{V,\rm SSP}$ [mag] & 0.11 & 0.43 & 0.23 & 1.00 & 0.60 & 0.22 & 0.00 & 0.12 & 0.00 & 0.18 & 0.71 & 0.06 \\
$\chi^2_{\rm DB}$/$\chi^2_{\rm SSP}$ & 1.01 & 1.03 & 1.27 & 1.00 & 1.05 & 0.95 & 0.72 & 0.92 & 0.94 & 0.93 & 0.51 & 2.02 \\
        \hline
\hline
    \end{tabular}
    \\
    \vspace{1ex}
    {\raggedright \footnotesize $^a$: Categories for the individual objects: GC: globular cluster candidates, C: possibly contaminated by galactic light (or by a nearby diffraction spike for id 9), E: extended sources from visual inspection, B: in the bulk of the galaxy or actively star forming, for e.g. the OIII regions in Figure 4. \par}
    {\raggedright \footnotesize $^b$: The stellar masses account for stellar mass loss but are not corrected for magnification factors, which are $\sim 10-100$ and can vary across the image. \par}
    \label{tab:sparkle_properties}
\end{table*}

\begin{acknowledgments}
ACKNOWLEDGEMENTS: 

This work is based in part on observations made with the NASA/ESA/CSA James Webb Space Telescope. The data were obtained from the Mikulski Archive for Space Telescopes at the Space Telescope Science Institute, which is operated by the Association of Universities for Research in Astronomy, Inc., under NASA contract NAS 5-03127 for JWST. The Early Release Observations and associated materials were developed, executed, and compiled by the ERO production team: Hannah Braun, Claire Blome, Matthew Brown, Margaret Carruthers, Dan Coe, Joseph DePasquale, Nestor Espinoza, Macarena Garcia Marin, Karl Gordon, Alaina Henry, Leah Hustak, Andi James, Ann Jenkins, Anton Koekemoer, Stephanie LaMassa, David Law, Alexandra Lockwood, Amaya Moro-Martin, Susan Mullally, Alyssa Pagan, Dani Player, Klaus Pontoppidan, Charles Proffitt, Christine Pulliam, Leah Ramsay, Swara Ravindranath, Neill Reid, Massimo Robberto, Elena Sabbi, Leonardo Ubeda. The EROs were also made possible by the foundational efforts and support from the JWST instruments, STScI planning and scheduling, and Data Management teams.

This research was supported by Natural Sciences and Engineering Research Council (NSERC) of Canada Discovery Grants to RA, AM, and MS, by an NSERC Discovery Accelerator to MS, and by a grant from the Canadian Space Agency to the NIRISS GTO Team (18JWST-GTO1). MB acknowledges support from the Slovenian national research agency ARRS through grant N1-0238.
Funding from the Dunlap Institute for Astronomy \& Astrophysics is gratefully acknowledged. The Dunlap Institute is funded through an endowment established by the David Dunlap family and the
University of Toronto. This research used the Canadian Advanced Network For Astronomy Research (CANFAR) operated in partnership by the Canadian Astronomy Data Centre and The Digital Research Alliance of Canada with support from the National Research Council of Canada, the Canadian Space Agency, CANARIE and the Canadian Foundation for Innovation.
\end{acknowledgments}

\vspace{5mm}
\facilities{JWST, HST(ACS)}


\software{\texttt{astropy} \citep{2013A&A...558A..33A,2018AJ....156..123A}, 
Photutils \citep{photutils21},
          \texttt{Cloudy} \citep{2013RMxAA..49..137F}, \texttt{SEP} (\citealt{Barbary2016}, \citealt{1996A&AS..117..393B}),
          \texttt{SExtractor} \citep{1996A&AS..117..393B},
          \texttt{FSPS} \citep{ben_johnson_2021_4737461},
          Dense Basis \citep{iyer21},
          \texttt{matplotlib} \citep{caswell2019matplotlib}, 
          \texttt{scipy} \citep{virtanen2020scipy}, 
          \texttt{numpy} \citep{walt2011numpy}, 
          \texttt{corner} \citep{Foreman-Mackey2016},
          \texttt{hickle} \citep{Price2018}
          and \texttt{GALFIT} \citep{Peng2010}
          }





\bibliography{sparkly_sparkle}{}

\begin{thebibliography}{}
\expandafter\ifx\csname natexlab\endcsname\relax\def\natexlab#1{#1}\fi
\providecommand{\url}[1]{\href{#1}{#1}}

\bibitem[{{Astropy Collaboration} {et~al.}(2013){Astropy Collaboration},
  {Robitaille}, {Tollerud}, {Greenfield}, {Droettboom}, {Bray}, {Aldcroft},
  {Davis}, {Ginsburg}, {Price-Whelan}, {Kerzendorf}, {Conley}, {Crighton},
  {Barbary}, {Muna}, {Ferguson}, {Grollier}, {Parikh}, {Nair}, {Unther},
  {Deil}, {Woillez}, {Conseil}, {Kramer}, {Turner}, {Singer}, {Fox}, {Weaver},
  {Zabalza}, {Edwards}, {Azalee Bostroem}, {Burke}, {Casey}, {Crawford},
  {Dencheva}, {Ely}, {Jenness}, {Labrie}, {Lim}, {Pierfederici}, {Pontzen},
  {Ptak}, {Refsdal}, {Servillat}, \& {Streicher}}]{2013A&A...558A..33A}
{Astropy Collaboration}, {Robitaille}, T.~P., {Tollerud}, E.~J., {et~al.} 2013,
  \aap, 558, A33

\bibitem[{{Astropy Collaboration} {et~al.}(2018){Astropy Collaboration},
  {Price-Whelan}, {Sip{\H{o}}cz}, {G{\"u}nther}, {Lim}, {Crawford}, {Conseil},
  {Shupe}, {Craig}, {Dencheva}, {Ginsburg}, {VanderPlas}, {Bradley},
  {P{\'e}rez-Su{\'a}rez}, {de Val-Borro}, {Aldcroft}, {Cruz}, {Robitaille},
  {Tollerud}, {Ardelean}, {Babej}, {Bach}, {Bachetti}, {Bakanov}, {Bamford},
  {Barentsen}, {Barmby}, {Baumbach}, {Berry}, {Biscani}, {Boquien}, {Bostroem},
  {Bouma}, {Brammer}, {Bray}, {Breytenbach}, {Buddelmeijer}, {Burke},
  {Calderone}, {Cano Rodr{\'\i}guez}, {Cara}, {Cardoso}, {Cheedella}, {Copin},
  {Corrales}, {Crichton}, {D'Avella}, {Deil}, {Depagne}, {Dietrich}, {Donath},
  {Droettboom}, {Earl}, {Erben}, {Fabbro}, {Ferreira}, {Finethy}, {Fox},
  {Garrison}, {Gibbons}, {Goldstein}, {Gommers}, {Greco}, {Greenfield},
  {Groener}, {Grollier}, {Hagen}, {Hirst}, {Homeier}, {Horton}, {Hosseinzadeh},
  {Hu}, {Hunkeler}, {Ivezi{\'c}}, {Jain}, {Jenness}, {Kanarek}, {Kendrew},
  {Kern}, {Kerzendorf}, {Khvalko}, {King}, {Kirkby}, {Kulkarni}, {Kumar},
  {Lee}, {Lenz}, {Littlefair}, {Ma}, {Macleod}, {Mastropietro}, {McCully},
  {Montagnac}, {Morris}, {Mueller}, {Mumford}, {Muna}, {Murphy}, {Nelson},
  {Nguyen}, {Ninan}, {N{\"o}the}, {Ogaz}, {Oh}, {Parejko}, {Parley}, {Pascual},
  {Patil}, {Patil}, {Plunkett}, {Prochaska}, {Rastogi}, {Reddy Janga},
  {Sabater}, {Sakurikar}, {Seifert}, {Sherbert}, {Sherwood-Taylor}, {Shih},
  {Sick}, {Silbiger}, {Singanamalla}, {Singer}, {Sladen}, {Sooley},
  {Sornarajah}, {Streicher}, {Teuben}, {Thomas}, {Tremblay}, {Turner},
  {Terr{\'o}n}, {van Kerkwijk}, {de la Vega}, {Watkins}, {Weaver}, {Whitmore},
  {Woillez}, {Zabalza}, \& {Astropy Contributors}}]{2018AJ....156..123A}
{Astropy Collaboration}, {Price-Whelan}, A.~M., {Sip{\H{o}}cz}, B.~M., {et~al.}
  2018, \aj, 156, 123

\bibitem[{Barbary(2016)}]{Barbary2016}
Barbary, K. 2016, J. Open Source Softw., 1, 58

\bibitem[{{Baumgardt}(2006)}]{2006astro.ph..5125B}
{Baumgardt}, H. 2006, arXiv e-prints, astro

\bibitem[{{Baumgardt} \& {Makino}(2003)}]{Baumgardt2003}
{Baumgardt}, H., \& {Makino}, J. 2003, \mnras, 340, 227

\bibitem[{{Bennet} {et~al.}(2022){Bennet}, {Alfaro-Cuello}, {del Pino},
  {Watkins}, {van der Marel}, \& {Sohn}}]{bennet22}
{Bennet}, P., {Alfaro-Cuello}, M., {del Pino}, A., {et~al.} 2022, arXiv
  e-prints, arXiv:2207.13100

\bibitem[{{Bertin} \& {Arnouts}(1996)}]{1996A&AS..117..393B}
{Bertin}, E., \& {Arnouts}, S. 1996, \aaps, 117, 393

\bibitem[{{Bradley} {et~al.}(2021){Bradley}, {Sip{\H{o}}cz}, {Robitaille},
  {Tollerud}, {Vin{\'\i}cius}, {Deil}, {Barbary}, {Wilson}, {Busko}, {Donath},
  {G{\"u}nther}, {Cara}, {Conseil}, {Bostroem}, {Droettboom}, {Bray},
  {Krachyon}, {Lim}, {Andersen Bratholm}, {Barentsen}, {Craig}, {Rathi},
  {Pascual}, {Perren}, {Georgiev}, {De Val-Borro}, {Kerzendorf}, {Bach},
  {Quint}, \& {Souchereau}}]{photutils21}
{Bradley}, L., {Sip{\H{o}}cz}, B., {Robitaille}, T., {et~al.} 2021,
  {astropy/photutils: 1.1.0}, Zenodo, v1.1.0,  Zenodo,
  doi:10.5281/zenodo.4624996

\bibitem[{{Brammer} \& {Matharu}(2021)}]{2021zndo...5012699B}
{Brammer}, G., \& {Matharu}, J. 2021, {gbrammer/grizli: Release 2021}, Zenodo,
  v1.3.2,  Zenodo, doi:10.5281/zenodo.5012699

\bibitem[{{Brammer} {et~al.}(2008){Brammer}, {van Dokkum}, \&
  {Coppi}}]{2008ApJ...686.1503B}
{Brammer}, G.~B., {van Dokkum}, P.~G., \& {Coppi}, P. 2008, \apj, 686, 1503

\bibitem[{{Brodie} \& {Strader}(2006)}]{2006ARA&A..44..193B}
{Brodie}, J.~P., \& {Strader}, J. 2006, \araa, 44, 193

\bibitem[{{Caminha} {et~al.}(2022){Caminha}, {Suyu}, {Mercurio}, {Brammer},
  {Bergamini}, {Vanzella}, \& {Acebron}}]{caminha22}
{Caminha}, G.~B., {Suyu}, S.~H., {Mercurio}, A., {et~al.} 2022, arXiv e-prints,
  arXiv:2207.07567

\bibitem[{{Carlberg}(2002)}]{carlberg2002}
{Carlberg}, R.~G. 2002, \apj, 573, 60

\bibitem[{Caswell {et~al.}(2019)Caswell, Droettboom, Hunter, Firing, Lee,
  Klymak, Stansby, de~Andrade, Nielsen, Varoquaux,
  {et~al.}}]{caswell2019matplotlib}
Caswell, T., Droettboom, M., Hunter, J., {et~al.} 2019, matplotlib/matplotlib
  v3. 1.0,  May

\bibitem[{{Coe} {et~al.}(2019){Coe}, {Salmon}, {Brada{\v{c}}}, {Bradley},
  {Sharon}, {Zitrin}, {Acebron}, {Cerny}, {Cibirka}, {Strait},
  {Paterno-Mahler}, {Mahler}, {Avila}, {Ogaz}, {Huang}, {Pelliccia}, {Stark},
  {Mainali}, {Oesch}, {Trenti}, {Carrasco}, {Dawson}, {Rodney}, {Strolger},
  {Riess}, {Jones}, {Frye}, {Czakon}, {Umetsu}, {Vulcani}, {Graur}, {Jha},
  {Graham}, {Molino}, {Nonino}, {Hjorth}, {Selsing}, {Christensen},
  {Kikuchihara}, {Ouchi}, {Oguri}, {Welch}, {Lemaux}, {Andrade-Santos}, {Hoag},
  {Johnson}, {Peterson}, {Past}, {Fox}, {Agulli}, {Livermore}, {Ryan}, {Lam},
  {Sendra-Server}, {Toft}, {Lovisari}, \& {Su}}]{Coe2019}
{Coe}, D., {Salmon}, B., {Brada{\v{c}}}, M., {et~al.} 2019, \apj, 884, 85

\bibitem[{{Conroy} {et~al.}(2009){Conroy}, {Gunn}, \& {White}}]{conroy09_fsps}
{Conroy}, C., {Gunn}, J.~E., \& {White}, M. 2009, \apj, 699, 486

\bibitem[{{Conroy} {et~al.}(2010){Conroy}, {White}, \& {Gunn}}]{conroy10_fsps}
{Conroy}, C., {White}, M., \& {Gunn}, J.~E. 2010, \apj, 708, 58

\bibitem[{{de Grijs} {et~al.}(2001){de Grijs}, {O'Connell}, \&
  {Gallagher}}]{deGrijs2001}
{de Grijs}, R., {O'Connell}, R.~W., \& {Gallagher}, John~S., I. 2001, \aj, 121,
  768

\bibitem[{{Doyon} {et~al.}(2012){Doyon}, {Hutchings}, {Beaulieu}, {Albert},
  {Lafreni{\`e}re}, {Willott}, {Touahri}, {Rowlands}, {Maszkiewicz},
  {Fullerton}, {Volk}, {Martel}, {Chayer}, {Sivaramakrishnan}, {Abraham},
  {Ferrarese}, {Jayawardhana}, {Johnstone}, {Meyer}, {Pipher}, \&
  {Sawicki}}]{doyon2012}
{Doyon}, R., {Hutchings}, J.~B., {Beaulieu}, M., {et~al.} 2012, in Society of
  Photo-Optical Instrumentation Engineers (SPIE) Conference Series, Vol. 8442,
  Space Telescopes and Instrumentation 2012: Optical, Infrared, and Millimeter
  Wave, ed. M.~C. {Clampin}, G.~G. {Fazio}, H.~A. {MacEwen}, \& J.~{Oschmann},
  Jacobus~M., 84422R

\bibitem[{{Erkal} {et~al.}(2019){Erkal}, {Belokurov}, {Laporte}, {Koposov},
  {Li}, {Grillmair}, {Kallivayalil}, {Price-Whelan}, {Evans}, {Hawkins},
  {Hendel}, {Mateu}, {Navarro}, {del Pino}, {Slater}, {Sohn}, \& {Orphan Aspen
  Treasury Collaboration}}]{erkal19}
{Erkal}, D., {Belokurov}, V., {Laporte}, C.~F.~P., {et~al.} 2019, \mnras, 487,
  2685

\bibitem[{{Ferland} {et~al.}(2013){Ferland}, {Porter}, {van Hoof}, {Williams},
  {Abel}, {Lykins}, {Shaw}, {Henney}, \& {Stancil}}]{2013RMxAA..49..137F}
{Ferland}, G.~J., {Porter}, R.~L., {van Hoof}, P.~A.~M., {et~al.} 2013, \rmxaa,
  49, 137

\bibitem[{{Forbes} {et~al.}(2018){Forbes}, {Bastian}, {Gieles}, {Crain},
  {Kruijssen}, {Larsen}, {Ploeckinger}, {Agertz}, {Trenti}, {Ferguson},
  {Pfeffer}, \& {Gnedin}}]{forbes2018}
{Forbes}, D.~A., {Bastian}, N., {Gieles}, M., {et~al.} 2018, Proceedings of the
  Royal Society of London Series A, 474, 20170616

\bibitem[{Foreman-Mackey(2016)}]{Foreman-Mackey2016}
Foreman-Mackey, D. 2016, Journal of Open Source Software, 1, 24.
\newblock \url{https://doi.org/10.21105/joss.00024}

\bibitem[{{F{\"o}rster Schreiber} {et~al.}(2006){F{\"o}rster Schreiber},
  {Genzel}, {Lehnert}, {Bouch{\'e}}, {Verma}, {Erb}, {Shapley}, {Steidel},
  {Davies}, {Lutz}, {Nesvadba}, {Tacconi}, {Eisenhauer}, {Abuter}, {Gilbert},
  {Gillessen}, \& {Sternberg}}]{Schreiber2006}
{F{\"o}rster Schreiber}, N.~M., {Genzel}, R., {Lehnert}, M.~D., {et~al.} 2006,
  \apj, 645, 1062

\bibitem[{{Freeman} \& {Norris}(1981)}]{1981ARA&A..19..319F}
{Freeman}, K.~C., \& {Norris}, J. 1981, \araa, 19, 319

\bibitem[{{Genzel} {et~al.}(2006){Genzel}, {Tacconi}, {Eisenhauer},
  {F{\"o}rster Schreiber}, {Cimatti}, {Daddi}, {Bouch{\'e}}, {Davies},
  {Lehnert}, {Lutz}, {Nesvadba}, {Verma}, {Abuter}, {Shapiro}, {Sternberg},
  {Renzini}, {Kong}, {Arimoto}, \& {Mignoli}}]{Genzel2006}
{Genzel}, R., {Tacconi}, L.~J., {Eisenhauer}, F., {et~al.} 2006, \nat, 442, 786

\bibitem[{Golubchik {et~al.}(2022)Golubchik, Furtak, Meena, \&
  Zitrin}]{golubchik22}
Golubchik, M., Furtak, L.~J., Meena, A.~K., \& Zitrin, A. 2022,
  arXiv:2207.05007 [astro-ph].
\newblock \url{http://arxiv.org/abs/2207.05007}

\bibitem[{{Harris} \& {Racine}(1979)}]{1979ARA&A..17..241H}
{Harris}, W.~E., \& {Racine}, R. 1979, \araa, 17, 241

\bibitem[{{Iyer} \& {Gawiser}(2017)}]{iyer17}
{Iyer}, K., \& {Gawiser}, E. 2017, \apj, 838, 127

\bibitem[{{Iyer} {et~al.}(2018){Iyer}, {Gawiser}, {Dav{\'e}}, {Davis},
  {Finkelstein}, {Kodra}, {Koekemoer}, {Kurczynski}, {Newman}, {Pacifici}, \&
  {Somerville}}]{iyer18}
{Iyer}, K., {Gawiser}, E., {Dav{\'e}}, R., {et~al.} 2018, \apj, 866, 120

\bibitem[{{Iyer} {et~al.}(2019){Iyer}, {Gawiser}, {Faber}, {Ferguson},
  {Kartaltepe}, {Koekemoer}, {Pacifici}, \& {Somerville}}]{iyer19}
{Iyer}, K.~G., {Gawiser}, E., {Faber}, S.~M., {et~al.} 2019, \apj, 879, 116

\bibitem[{{Iyer} {et~al.}(2021){Iyer}, {Gawiser}, {Faber}, {Ferguson},
  {Kartaltepe}, {Koekemoer}, {Pacifici}, \& {Somerville}}]{iyer21}
---. 2021, {dense\_basis: Dense Basis SED fitting}, Astrophysics Source Code
  Library, record ascl:2104.015, , , ascl:2104.015

\bibitem[{Johnson {et~al.}(2021)Johnson, Foreman-Mackey, Sick, Leja, Byler,
  Walmsley, Tollerud, Leung, \& Scott}]{ben_johnson_2021_4737461}
Johnson, B., Foreman-Mackey, D., Sick, J., {et~al.} 2021, dfm/python-fsps:
  python-fsps v0.4.1rc1, vv0.4.1rc1,  Zenodo, doi:10.5281/zenodo.4737461.
\newblock \url{https://doi.org/10.5281/zenodo.4737461}

\bibitem[{{Johnson} {et~al.}(2017){Johnson}, {Caldwell}, {Rich}, {Mateo},
  {Bailey}, {Clarkson}, {Olszewski}, \& {Walker}}]{johnson17}
{Johnson}, C.~I., {Caldwell}, N., {Rich}, R.~M., {et~al.} 2017, \apj, 836, 168

\bibitem[{{Kroupa}(2001)}]{Kroupa2001}
{Kroupa}, P. 2001, \mnras, 322, 231

\bibitem[{{Leja} {et~al.}(2019){Leja}, {Carnall}, {Johnson}, {Conroy}, \&
  {Speagle}}]{leja19}
{Leja}, J., {Carnall}, A.~C., {Johnson}, B.~D., {Conroy}, C., \& {Speagle},
  J.~S. 2019, \apj, 876, 3

\bibitem[{{Livermore} {et~al.}(2012){Livermore}, {Jones}, {Richard}, {Bower},
  {Ellis}, {Swinbank}, {Rigby}, {Smail}, {Arribas}, {Rodriguez Zaurin},
  {Colina}, {Ebeling}, \& {Crain}}]{livermore12}
{Livermore}, R.~C., {Jones}, T., {Richard}, J., {et~al.} 2012, \mnras, 427, 688

\bibitem[{{Livermore} {et~al.}(2015){Livermore}, {Jones}, {Richard}, {Bower},
  {Swinbank}, {Yuan}, {Edge}, {Ellis}, {Kewley}, {Smail}, {Coppin}, \&
  {Ebeling}}]{livermore15}
{Livermore}, R.~C., {Jones}, T.~A., {Richard}, J., {et~al.} 2015, \mnras, 450,
  1812

\bibitem[{{Lower} {et~al.}(2020){Lower}, {Narayanan}, {Leja}, {Johnson},
  {Conroy}, \& {Dav{\'e}}}]{lower20}
{Lower}, S., {Narayanan}, D., {Leja}, J., {et~al.} 2020, \apj, 904, 33

\bibitem[{{Mahler} {et~al.}(2022){Mahler}, {Jauzac}, {Richard}, {Beauchesne},
  {Ebeling}, {Lagattuta}, {Natarajan}, {Sharon}, {Atek}, {Claeyssens},
  {Cl{\'e}ment}, {Eckert}, {Edge}, {Kneib}, \& {Niemiec}}]{mahler22}
{Mahler}, G., {Jauzac}, M., {Richard}, J., {et~al.} 2022, arXiv e-prints,
  arXiv:2207.07101

\bibitem[{{Ocvirk} {et~al.}(2006){Ocvirk}, {Pichon}, {Lan{\c{c}}on}, \&
  {Thi{\'e}baut}}]{ocvirk06}
{Ocvirk}, P., {Pichon}, C., {Lan{\c{c}}on}, A., \& {Thi{\'e}baut}, E. 2006,
  \mnras, 365, 46

\bibitem[{{Olsen} {et~al.}(2021){Olsen}, {Gawiser}, {Iyer}, {McQuinn},
  {Johnson}, {Telford}, {Wright}, {Broussard}, \& {Kurczynski}}]{olsen21}
{Olsen}, C., {Gawiser}, E., {Iyer}, K., {et~al.} 2021, \apj, 913, 45

\bibitem[{{Peebles} \& {Dicke}(1968)}]{1968ApJ...154..891P}
{Peebles}, P.~J.~E., \& {Dicke}, R.~H. 1968, \apj, 154, 891

\bibitem[{{Peng} {et~al.}(2010){Peng}, {Ho}, {Impey}, \& {Rix}}]{Peng2010}
{Peng}, C.~Y., {Ho}, L.~C., {Impey}, C.~D., \& {Rix}, H.-W. 2010, \aj, 139,
  2097

\bibitem[{{Pontoppidan} {et~al.}(2022){Pontoppidan}, {Blome}, {Braun}, {Brown},
  {Carruthers}, {Coe}, {DePasquale}, {Espinoza}, {Garcia Marin}, {Gordon},
  {Henry}, {Hustak}, {James}, {Koekemoer}, {LaMassa}, {Law}, {Lockwood},
  {Moro-Martin}, {Mullally}, {Pagan}, {Player}, {Proffitt}, {Pulliam},
  {Ramsay}, {Ravindranath}, {Reid}, {Robberto}, {Sabbi}, \&
  {Ubeda}}]{2022arXiv220713067P}
{Pontoppidan}, K., {Blome}, C., {Braun}, H., {et~al.} 2022, arXiv e-prints,
  arXiv:2207.13067

\bibitem[{Price {et~al.}(2018)Price, van~der Velden, Celles, Eendebak, McKerns,
  Olson, Raffel, Yi, \& Ash}]{Price2018}
Price, D.~C., van~der Velden, E., Celles, S., {et~al.} 2018, Journal of Open
  Source Software, 3, 1115.
\newblock \url{https://doi.org/10.21105/joss.01115}

\bibitem[{{Renzini}(2017)}]{renzini2017}
{Renzini}, A. 2017, \mnras, 469, L63

\bibitem[{{Rieke} {et~al.}(2005){Rieke}, {Kelly}, \& {Horner}}]{rieke2005}
{Rieke}, M.~J., {Kelly}, D., \& {Horner}, S. 2005, in Society of Photo-Optical
  Instrumentation Engineers (SPIE) Conference Series, Vol. 5904, Cryogenic
  Optical Systems and Instruments XI, ed. J.~B. {Heaney} \& L.~G. {Burriesci},
  1--8

\bibitem[{{Schweizer} \& {Seitzer}(1998)}]{Schweizer98}
{Schweizer}, F., \& {Seitzer}, P. 1998, \aj, 116, 2206

\bibitem[{{Trujillo-Gomez} {et~al.}(2021){Trujillo-Gomez}, {Kruijssen},
  {Reina-Campos}, {Pfeffer}, {Keller}, {Crain}, {Bastian}, \&
  {Hughes}}]{TrujilloGomez2021}
{Trujillo-Gomez}, S., {Kruijssen}, J.~M.~D., {Reina-Campos}, M., {et~al.} 2021,
  \mnras, 503, 31

\bibitem[{{Vanzella} {et~al.}(2017){Vanzella}, {Calura}, {Meneghetti},
  {Mercurio}, {Castellano}, {Caminha}, {Balestra}, {Rosati}, {Tozzi}, {De
  Barros}, {Grazian}, {D'Ercole}, {Ciotti}, {Caputi}, {Grillo}, {Merlin},
  {Pentericci}, {Fontana}, {Cristiani}, \& {Coe}}]{vanzella17}
{Vanzella}, E., {Calura}, F., {Meneghetti}, M., {et~al.} 2017, \mnras, 467,
  4304

\bibitem[{{Vanzella} {et~al.}(2022){Vanzella}, {Castellano}, {Bergamini},
  {Treu}, {Mercurio}, {Scarlata}, {Rosati}, {Grillo}, {Acebron}, {Caminha},
  {Nonino}, {Nanayakkara}, {Roberts-Borsani}, {Bradac}, {Wang}, {Brammer},
  {Strait}, {Vulcani}, {Mestric}, {Meneghetti}, {Calura}, {Henry}, {Zanella},
  {Trenti}, {Boyett}, {Morishita}, {Calabro}, {Glazebrook}, {Marchesini},
  {Birrer}, {Yang}, \& {Jones}}]{vanzella2022}
{Vanzella}, E., {Castellano}, M., {Bergamini}, P., {et~al.} 2022, arXiv
  e-prints, arXiv:2208.00520

\bibitem[{Virtanen {et~al.}(2020)Virtanen, Gommers, Oliphant, Haberland, Reddy,
  Cournapeau, Burovski, Peterson, Weckesser, Bright,
  {et~al.}}]{virtanen2020scipy}
Virtanen, P., Gommers, R., Oliphant, T.~E., {et~al.} 2020, Nature methods, 1

\bibitem[{Walt {et~al.}(2011)Walt, Colbert, \& Varoquaux}]{walt2011numpy}
Walt, S. v.~d., Colbert, S.~C., \& Varoquaux, G. 2011, Computing in Science \&
  Engineering, 13, 22

\bibitem[{{Weaver} {et~al.}(2022){Weaver}, {Kauffmann}, {Ilbert}, {McCracken},
  {Moneti}, {Toft}, {Brammer}, {Shuntov}, {Davidzon}, {Hsieh}, {Laigle},
  {Anastasiou}, {Jespersen}, {Vinther}, {Capak}, {Casey}, {McPartland},
  {Milvang-Jensen}, {Mobasher}, {Sanders}, {Zalesky}, {Arnouts}, {Aussel},
  {Dunlop}, {Faisst}, {Franx}, {Furtak}, {Fynbo}, {Gould}, {Greve}, {Gwyn},
  {Kartaltepe}, {Kashino}, {Koekemoer}, {Kokorev}, {Le F{\`e}vre}, {Lilly},
  {Masters}, {Magdis}, {Mehta}, {Peng}, {Riechers}, {Salvato}, {Sawicki},
  {Scarlata}, {Scoville}, {Shirley}, {Silverman}, {Sneppen}, {Smolc̆i{\'c}},
  {Steinhardt}, {Stern}, {Tanaka}, {Taniguchi}, {Teplitz}, {Vaccari}, {Wang},
  \& {Zamorani}}]{weaver22}
{Weaver}, J.~R., {Kauffmann}, O.~B., {Ilbert}, O., {et~al.} 2022, \apjs, 258,
  11

\bibitem[{{Welch} {et~al.}(2022{\natexlab{a}}){Welch}, {Coe}, {Zitrin},
  {Diego}, {Windhorst}, {Mandelker}, {Vanzella}, {Ravindranath}, {Zackrisson},
  {Florian}, {Bradley}, {Sharon}, {Brada{\v{c}}}, {Rigby}, {Frye}, \&
  {Fujimoto}}]{welch2022}
{Welch}, B., {Coe}, D., {Zitrin}, A., {et~al.} 2022{\natexlab{a}}, arXiv
  e-prints, arXiv:2207.03532

\bibitem[{{Welch} {et~al.}(2022{\natexlab{b}}){Welch}, {Coe}, {Diego},
  {Zitrin}, {Zackrisson}, {Dimauro}, {Jim{\'e}nez-Teja}, {Kelly}, {Mahler},
  {Oguri}, {Timmes}, {Windhorst}, {Florian}, {de Mink}, {Avila}, {Anderson},
  {Bradley}, {Sharon}, {Vikaeus}, {McCandliss}, {Brada{\v{c}}}, {Rigby},
  {Frye}, {Toft}, {Strait}, {Trenti}, {Sharma}, {Andrade-Santos}, \&
  {Broadhurst}}]{welch22}
{Welch}, B., {Coe}, D., {Diego}, J.~M., {et~al.} 2022{\natexlab{b}}, \nat, 603,
  815

\bibitem[{{Willott} {et~al.}(2022){Willott}, {Doyon}, {Albert}, {Brammer},
  {Dixon}, {Muzic}, {Ravindranath}, {Scholz}, {Abraham}, {Artigau},
  {Brada{\v{c}}}, {Goudfrooij}, {Hutchings}, {Iyer}, {Jayawardhana}, {LaMassa},
  {Martis}, {Meyer}, {Morishita}, {Mowla}, {Muzzin}, {Noirot}, {Pacifici},
  {Rowlands}, {Sarrouh}, {Sawicki}, {Taylor}, {Volk}, \& {Zabl}}]{willott22}
{Willott}, C.~J., {Doyon}, R., {Albert}, L., {et~al.} 2022, \pasp, 134, 025002

\bibitem[{{Wuyts} {et~al.}(2014){Wuyts}, {Rigby}, {Gladders}, \&
  {Sharon}}]{wuyts14}
{Wuyts}, E., {Rigby}, J.~R., {Gladders}, M.~D., \& {Sharon}, K. 2014, \apj,
  781, 61

\end{thebibliography}
\bibliographystyle{aasjournal}

\end{document}